\documentclass[conference]{IEEEtran}
\usepackage{cite}
\usepackage{amsmath,amssymb,amsfonts}
\usepackage{algorithmic}
\usepackage{graphicx,dblfloatfix}
\usepackage{textcomp}
\usepackage{float}
\usepackage{subfigure}
\usepackage{listings}
\usepackage{xcolor}
\usepackage{enumitem}
\def\BibTeX{{\rm B\kern-.05em{\sc i\kern-.025em b}\kern-.08em
    T\kern-.1667em\lower.7ex\hbox{E}\kern-.125emX}}

\definecolor{codegreen}{rgb}{0,0.6,0}
\definecolor{codegray}{rgb}{0.5,0.5,0.5}
\definecolor{codepurple}{rgb}{0.58,0,0.82}
\definecolor{backcolour}{rgb}{0.95,0.95,0.92}
 
\lstdefinestyle{mystyle}{
    backgroundcolor=\color{backcolour},   
    commentstyle=\color{magenta},
    keywordstyle=\bfseries\color{codegreen},
    numberstyle=\tiny\color{codegray},
    stringstyle=\color{codepurple},
    basicstyle=\footnotesize\ttfamily\bfseries,
    breakatwhitespace=false,         
    breaklines=true,                 
    captionpos=b,                    
    keepspaces=true,                 
    numbers=left,                    
    numbersep=5pt,                  
    showspaces=false,                
    showstringspaces=false,
    showtabs=false,
    showlines=true,
    tabsize=2
}
 
\begin{document}

\title{AdaptMemBench: Application-Specific Memory Subsystem Benchmarking}

\author{\IEEEauthorblockN{Mahesh Lakshminarasimhan}
\IEEEauthorblockA{\textit{Department of Computer Science} \\
\textit{Boise State University}\\
Boise, Idaho, USA \\
maheshlakshminar@boisestate.edu}
\and
\IEEEauthorblockN{Catherine Olschanowsky}
\IEEEauthorblockA{\textit{Department of Computer Science} \\
\textit{Boise State University}\\
Boise, Idaho, USA \\
catherineolschan@boisestate.edu}
}

\maketitle
\thispagestyle{plain}
\pagestyle{plain}

\begin{abstract}
Optimizing scientific applications to take full advantage of modern memory subsystems is a continual challenge for application and compiler developers. 
Factors beyond working set size affect performance. 
A benchmark framework that explores the performance in an 
application-specific manner is essential to characterize memory performance and at the same time inform memory-efficient coding practices. 
We present AdaptMemBench, a configurable benchmark framework that measures achieved memory performance by emulating application-specific access patterns with a set of kernel-independent driver templates. 
This framework can explore the performance characteristics of a wide range of access patterns and can be used as a testbed for potential optimizations due to the flexibility of polyhedral code generation. 
We demonstrate the effectiveness of AdaptMemBench with case studies on commonly used computational kernels such as triad and multidimensional stencil patterns. 

\end{abstract}

\begin{IEEEkeywords}
Benchmarking, Memory Performance, Code Generation, Stencil Computations, Tiling Optimizations.
\end{IEEEkeywords}

\section{Introduction}

Scientific application performance is a function of memory bandwidth, instruction mix and order, memory footprint, and memory access patterns. The contribution of each is often not clear and interdependencies exist between each variable. This complexity, combined with the difficulty of instrumenting large application makes efficient optimization of these applications difficult. AdaptMemBench provides a framework for application developers and optimization experts to isolate portions of their application and measure execution characteristics. The framework provides a starting point to identify performance bottlenecks, identify potential optimizations, and explore the potential gains of those optimizations. 

Application performance is often bottlenecked by interaction with the memory subsystem due to the memory wall~\cite{wulf1995hitting}. Modern architectures combat this by using deep memory hierarchies and physically fragmented system memory. Reducing working set sizes is considered a good first step in optimization to take advantage of the caching capability of machines. However, optimizing is more complex than that, especially when dealing with shared memory parallelization. Memory access patterns, instruction mix, data sharing across caches, and vectorizability must all be considered in concert.

Selecting and applying optimizations remains a primary challenge during performance enhancements.  
Testing and understanding optimizations in situ when working with a large application can be cumbersome and error prone.
Given the difficulty around manipulating access patterns in situ, fewer optimization strategies are attempted and potential performance improvements are overlooked. 
Additionally, performance tools such as hardware counters, remain difficult to use in the context of a large application. The combination of these factors discourages effective optimizations.

A framework that allows extracted code to be isolated and measured will benefit the optimization process for specific projects, and will improve the reliability and reproducibility of performance experiments in the compiler optimization and programming construct research communities. During the exploration and experimentation phase, many different variants of the same code are produced. Tracking the differences between variants and maintaining correct execution becomes time consuming and challenging. A shared framework that supports experimentation and tracks code versions while outputting metadata with measurements will ease this challenge.

We propose a tool to explore the design landscape of the target architecture. 
The \emph{AdaptMemBench} framework can be used to measure system performance and to guide application-specific optimization decisions.
Expensive kernels extracted from larger applications can be manipulated in isolation to find the best optimization strategies. 
The framework reduces the amount of code that is transferred and provides mechanisms to experiment with 
data storage layout, execution order, and parallelization strategies.

AdaptMemBench provides several execution templates. The templates are combined with user provided code segments. The templates provide a common command line interface, handle all timing and hardware counter code, and output metadata and measurements in a common format. The code segments provided by the user can be expressed as C code or by using the polyhedral model. The latter provides a convenient mechanism for optimization experiments.

Several benchmarks \cite{mccalpin1995stream,kamil2005impact,snavely2002framework,smith2008bandwidth,strohmaier2005apexlatest,bailey1991parallel,luszczek2005introduction} exist that measure machine performance, with the benchmarking results conveying essential information about the application performance on the memory hierarchy of the machine. Existing memory benchmarks \cite{mccalpin1995stream,snavely2002framework,smith2008bandwidth} measure performance using a limited collection of streaming access patterns. However, benchmarking application-specific patterns that tend to be more complex remains a challenge. Current benchmarks \cite{mccalpin1995survey,kamil2005impact} are further constrained by the data sizes which can be executed, specifically in the higher levels of the memory subsystem. 

AdaptMemBench differs from previous efforts by incorporating polyhedral code generation.
This creates a configurable benchmarking framework that measures achieved memory bandwidth while mimicking application-specific memory access patterns. 
The Polyhedral model~\cite{verdoolaege2012polyhedral} simplifies writing the initial benchmark and provides a mechanism
to automatically transform the code.
Furthermore, our benchmark supports parallel applications and systems, and measures memory performance for data sizes across all levels of the memory hierarchy.

The primary contribution of this paper is a description and various demonstrations of the AdaptMemBench framework. 
Additionally, the framework was used to explore the performance of our university's HPC cluster.
The contributions of this paper include the following:
\begin{itemize}
\item A configurable benchmarking framework for application-specific memory performance characterization.
\item A detailed performance study on common computational kernels found in scientific applications for the impact of implicit locks, shared data spaces and false sharing.
\item An interleaved optimization strategy and demonstrated effectiveness for the triad pattern.
\item An evaluation of the efficacy of spatial tiling strategies for multidimensional Jacobi patterns using AdaptMemBench.

\end{itemize}

\begin{figure}[t]
  \includegraphics[width=1.01\columnwidth]{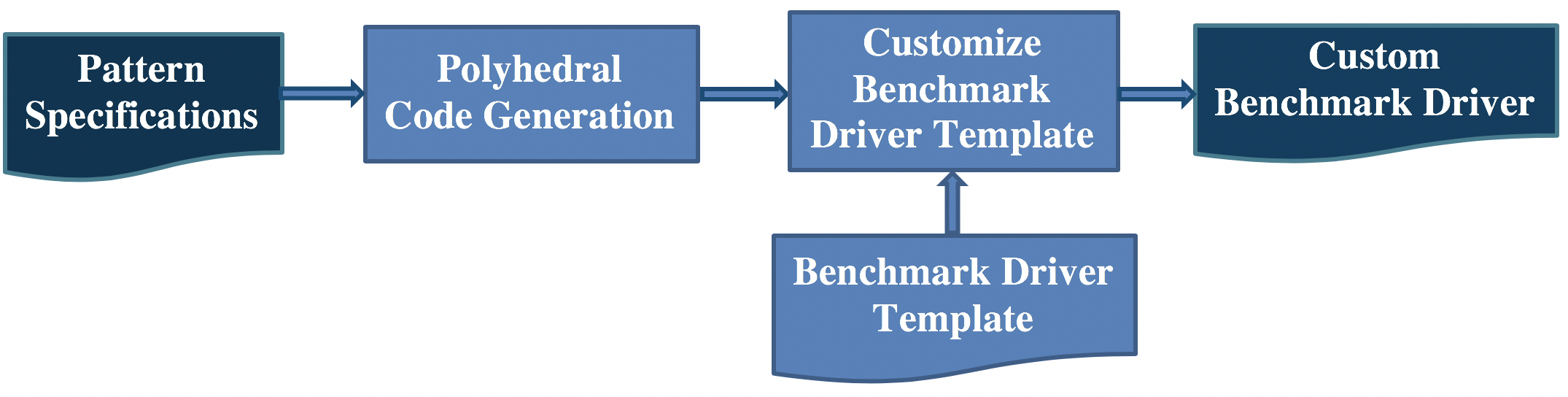}
  \caption{The proposed framework. \label{fig:workflow}}
\end{figure}

\section{AdaptMemBench Design}
The AdaptMemBench framework separates the user interface, validation, and output of the benchmark from the
code being measured and provide low overhead access to PAPI.
Figure~\ref{fig:workflow} illustrates the building blocks of the framework. 
Each computational kernel of interest is coded in a pattern specification.
If that pattern specification involves the polyhedral model it is passed through a polyhedral compiler.
The resulting (or original) c code is compiled together with one of several potential templates.
The templates provide a uniform interface and handle code to vary the working set size to cover each portion of the memory hierarchy,
along with timing, PAPI data collection, and output formatting.
The use of the polyhedral model adds a great deal of flexibility in terms of exploring optimizations.
The following subsection provides a brief overview of the polyhedral model. After the overview, the benchmark framework is described.

\subsection{Polyhedral Code Generation}
\begin{figure}[htb]
  \centering
  \includegraphics[width=1.0\columnwidth]{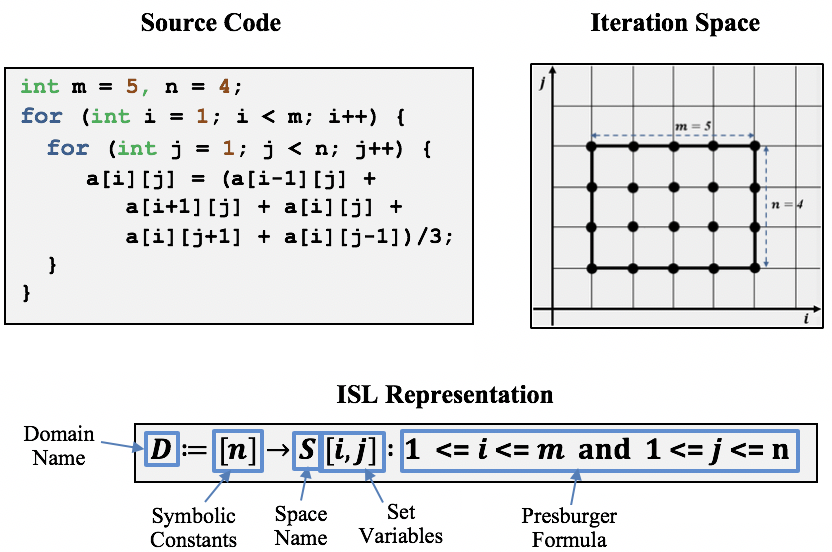}
  \caption{An example of polyhedral code generation with ISCC/ISL. \label{fig:codegen-gen}}
\end{figure}

Polyhedral code generation enables loop constructs to be expressed and manipulated mathematically. 
The iteration sets can be expressed without ordering unless a specific ordering is required. 
Figure~\ref{fig:codegen-gen} shows a loop nest for solving the heat equation. 
The associated iteration space is shown graphically as a two-dimensional space $(i,j)$.
Each node in the graph represents an iteration.
The Presburger formula for this example is shown at the bottom of the figure.

Code generation is performed on sets through polyhedral scanning, the result is control flow that produces the 
iterations in lexicographical ordering. As expressed in Figure~\ref{fig:codegen-gen} the original code would be
produced.
Transformations on the code are realized through the application of relations (or functions).

\begin{figure}[htb]
  \centering
  \includegraphics[width=1.02\columnwidth]{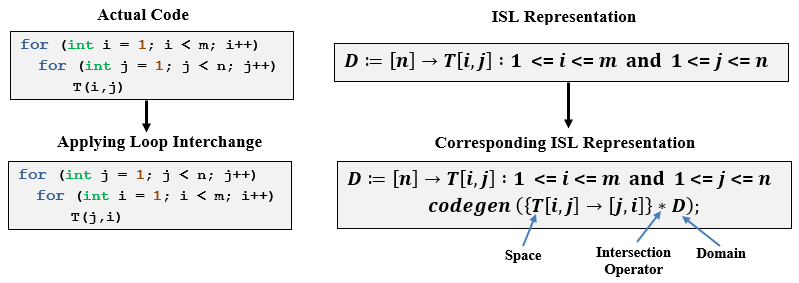}
  \caption{An illustration of loop interchange using ISCC. \label{fig:loop-interchange}}
\end{figure}

Loop interchange is a loop transformation that switches the order of two loops.
Figure~\ref{fig:loop-interchange} shows the relation used to apply loop interchange for the code in Figure~\ref{fig:codegen-gen}. For the relation from \texttt{\{i,j\}} to \texttt{\{j,i\}}, we apply the transformation on the execution domain defined, using the intersection operator. More complex transformations such as tiling can be performed with ease using the polyhedral model.

The polyhedral model represents iteration spaces that are affine. 
A significant amount of work has been done to expand the iteration spaces and schedules that can be represented, including work
that uses schedule trees for code generation within ISL~\cite{verdoolaege2010isl}. 
The Omega+ code generation tool is also able to incorporate iteration bounds based on runtime information using 
uninterpreted functions~\cite{kelly1998optimization}. Even with recent advances, the polyhedral model cannot express all
C kernels, and is, therefore, an optional step in the benchmark specification.

In the proposed benchmark, to automatically generate schedules for the application kernel initialization, execution, and validation, the ISCC \cite{verdoolaege2012polyhedral} polyhedral code generation tool is used, which offers an interface to the functionality provided by Integer Set Library (ISL)~\cite{verdoolaege2010isl} and Barvinok library~\cite{verdoolaege2007barvinok}. This tool enables the end user to manipulate sets and relations and generate source code reflecting their input.

\begin{figure}[tbh]
\begin{lstlisting}[firstnumber=1,label={lst:unified_DS},caption={The inner-most section of the Unified Data Spaces Template.},xleftmargin=2em,frame=single,framexleftmargin=1.5em,language=C]{Name}
...
//Execution
for(int k = 0; k < ntimes; k++) {
    #pragma omp parallel for CLAUSE
    #include "<kernel>_run.c"
}
...
\end{lstlisting}
\end{figure}
\begin{figure}[tbh]
\begin{lstlisting}[firstnumber=1,label={lst:ind_DS},caption={The inner-most section of the Independent Data Spaces Template.},xleftmargin=2em,frame=single,framexleftmargin=1.5em,language=C]{Name}
...
//Execution
#pragma omp parallel
{
  int t_id = omp_get_thread_num();
  for(int k = 0; k < ntimes; k++) {
      #include "<kernel>_run.c"
  }
}
...
\end{lstlisting}
\end{figure}

\subsection{Benchmark Implementation}
The proposed framework uses a set of generic benchmark driver templates for all variations of the access patterns. These driver templates provide a standard command line interface and a standard machine parsable and human readable output.  Currently, the framework supports the following three varieties of benchmark driver templates for shared memory applications:
\begin{enumerate}
    \item \textbf{The \textit{Unified Data Spaces} Template}  (Listing~\ref{lst:unified_DS}): The standard benchmarking template that utilizes unified data spaces shared among threads. It uses the work sharing and scheduling constructs offered by OpenMP to distribute resources among threads. The OpenMP clauses can be easily configured using the framework.

    \item \textbf{The \textit{Independent Data Spaces} Template} (Listing~\ref{lst:ind_DS}): This is a modified version of the unified data spaces template. It supports distinct data spaces separated into different memory regions accessed without any overlap, avoiding false sharing. As indicated by the experimental results that follow, benchmarking in this paradigm, yields optimal performance in the higher cache levels.

    \item \textbf{The \textit{PAPI Measurement} Template:} This template is built on top of the above two templates, using PAPI's low level API. The user is given an option to choose between the above two benchmarking paradigms and input the PAPI events to be recorded.
\end{enumerate}

Input pattern specifications consist of a header file and a set of ISCC input files. The initial step is to run the polyhedral code generator for the ISCC input files and transform them into corresponding C code files. The user-chosen driver template is then updated with the appropriate header and source files to create the customized \texttt{.cpp} benchmark driver code file. This benchmark driver code is compiled and executed with runtime arguments such as working set size, thread count and other parameters depending on the access pattern for which the benchmark is run.

\begin{figure}[tbh] 
  \includegraphics[width=1\columnwidth]{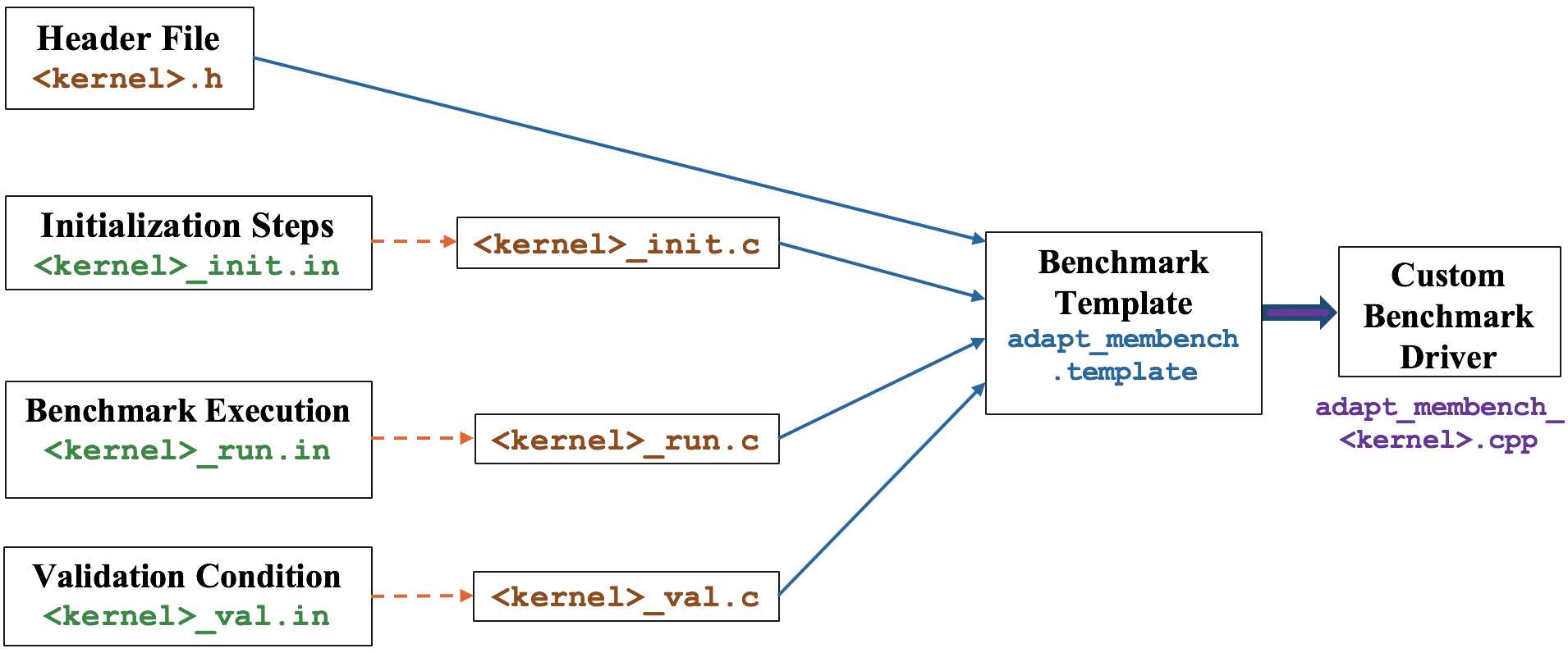}
  \caption{Implementation of the benchmark. \label{fig:implementation}}
\end{figure}

The purpose and functionality of each component in the pattern specifications shown in Figure~\ref{fig:implementation} are described below:
\begin{enumerate}
\item \textbf{Header file} (\texttt{<kernel>.h}): \\ This file contains the definitions of the memory mappings, statement macros, and the allocation code.
\begin{itemize}
\item \textit{Memory Mapping}: Indicates how the statements should map into memory using iterators as input.
\item \textit{Statement Macros}: The definition of the statement macros substituted in each of the C code files generated from the ISCC input files. Any data referred to within the statement should be referred to indirectly through the data mapping.
\item \textit{Allocation Code}: Specifies memory allocation of the data spaces used in the given application kernel. 
\end{itemize}
\item \textbf{Initialization steps} (\texttt{<kernel>\_init.in}):\\ This ISCC input file specifies the schedule for which the data domains allocated in the header file are initialized. In the C code file generated with ISCC, the associated statement macro specifying initialization steps is substituted when the benchmark is executed.
\item \textbf{Execution Schedule} (\texttt{<kernel>\_run.in}):\\ An ISCC input file that defines the iteration space in which the access pattern is executed. The application kernel defined as a macro in the header file is replaced in the \texttt{.c} file generated.  This code file consists of the for loop constructs associated with the execution domain which will be substituted in the driver when executed. 
\item \textbf{Validation condition} (\texttt{<kernel>\_val.in}):\\ This ISCC input file describes the schedule for which the results after executing the kernel is validated. The corresponding C code file generated is then called in the header file to validate the results.
\end{enumerate}

\section{Case Studies}
The performance characteristics of a set of computational kernels commonly used in 
performance studies are presented in this section.
The kernels are STREAM's triad and Jacobi 1D, 2D, and 3D. 
The kernels were chosen for their simplicity and well understood performance behaviors. 
The use cases demonstrate the need to separate implementation concerns when studying the performance of even simple 
kernels. The structure provided by AdaptMemBench improves the breadth of data collected and makes experiment reliability 
and reproducibility more easily attained. For each kernel we explore the impact of implicit locks, shared data spaces, 
and false sharing in SMP systems.

\noindent\textbf{Hardware:} Experiments were run on one of the nodes in the R2 HPC cluster at Boise State University, which has a 2.40GHz dual Intel Xeon E5-2680 v4 CPU. This node consists of two NUMA domains each containing 14 cores. Each core has a dedicated 32K L1 data cache and 256K L2 cache. The 35 MB L3 cache is shared among all the cores in each NUMA domain. The size of each cache line in this architecture is 64 bytes.

\noindent\textbf{Compilers:} GNU's gcc (version 6.3). When building C++ benchmark drivers, \texttt{-fopenmp} and \texttt{-O3} optimization flags were used. The \texttt{-lpapi} flag was set for PAPI-enabled benchmark drivers.

\noindent\textbf{Profiling Tool:} The benchmark drivers are instrumented with the Performance API (PAPI) \cite{mucci1999papi} library to access performance counters across the CPUs evaluated. PAPI is used to measure cache hits and the requests for exclusive access to cache lines.
 
\noindent\textbf{Problem size:} We executed the benchmarks with problem sizes across all levels of cache and those which exceeded the last-level cache and fit into the main memory. Each benchmark is executed for 1000 time iterations. The number of 
repetitions is configurable.

\subsection{The Triad benchmark}

We demonstrate the simplicity of AdaptMemBench by implementing the triad kernel from the STREAM benchmark due to its brevity and well-known performance. Listings \ref{lst:triad_header} and \ref{lst:triad_schedule}, along with the templates in Listings \ref{lst:unified_DS} and \ref{lst:ind_DS}  illustrate the process of creating a custom benchmark using a combination of input C code files, bypassing the polyhedral code generator. Alternatively, the kernel could have been expressed as a set: $\{[j]|0 <=j < n\}$.
The results are equivalent.

\begin{figure}[tbh]
\begin{lstlisting}[firstnumber=1,caption={Header file \texttt{<triad.h>}  for the triad benchmark.},label={lst:triad_header},xleftmargin=2em,frame=single,framexleftmargin=1.5em,language=C]{Name}
//Allocation Code
#define Triad_alloc double* A = double *) malloc(sizeof(double) * n); \
                    double* B = double *) malloc(sizeof(double) * n); \
                    double* C = double *) malloc(sizeof(double) * n);
//Memory Mapping
#define A_map(i) A[i]
#define B_map(i) B[i]
#define C_map(i) C[i]
//Initialization
#define Triad_init(i) A_map(i) = 1.0; B_map(i) = 3.0; C_map(i) = 4.0;
//Statement Definition
#define Triad_run(i) A_map(i) = B_map(i) + scalar * C_map(i);
//OpenMP clause
#define CLAUSE schedule(static)
\end{lstlisting}
\end{figure}

\begin{figure}
\begin{lstlisting}[firstnumber=1,caption={The execution schedule of the benchmark driver generated by combining the input file \texttt{<triad\_run.c>} and the template.}, label={lst:triad_schedule},xleftmargin=2em,frame=single,framexleftmargin=1.5em,language=C]{Name}
for (int j = 0; j < n; j++){
       Triad_run(j);   
}
\end{lstlisting}
\end{figure}
\begin{figure}[tbh]
  \centering
  \includegraphics[width=1\columnwidth]{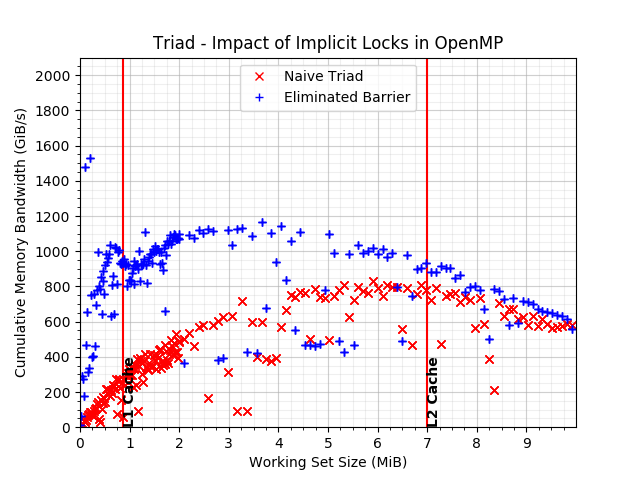}
  \caption{The impact of OpenMP barriers on achieved memory bandwidth. \label{fig:omp_barrier}}
\end{figure}

\subsection*{Cost of Barriers in OpenMP}
We use the triad benchmark generated to evaluate the overhead associated with barriers in OpenMP by using the \texttt{nowait} clause. With the AdaptMemBench framework, all that is required is to modify the definition of the macro \texttt{CLAUSE} to be \texttt{nowait}. 
As memory bandwidth results in Figure \ref{fig:omp_barrier} indicate, there is significant overhead caused by the barrier, and by breaking the barrier using the \texttt{nowait} clause we are able to achieve a reasonable speedup.
Though this modification may not be possible for all computations, e.g. those that have loop carried dependencies, our intention is just to demonstrate the performance degradation caused by compiler-induced locks using the simple triad kernel.


\begin{figure}[tbh]
\begin{lstlisting}[firstnumber=1,caption={Utilizing the OpenMP work sharing construct for data spaces of size \texttt{n} and \texttt{t} number of threads.},label={lst:omp_schedule},xleftmargin=2em,frame=single,framexleftmargin=1.5em,language=C]{Name}
for(int k = 0; k < ntimes; k++) {
    #pragma omp parallel for\
        schedule(static, n/t) nowait
    for (int i = 0; i < n; i++){
       A[i] = B[i] + scalar * C[i];
    }
}
\end{lstlisting}
\end{figure}
\begin{figure}[tbh]
\begin{lstlisting}[firstnumber=1,caption={The resultant triad benchmark using the \textit{independent data spaces} driver template},label={lst:independent_DS},xleftmargin=2em,frame=single,framexleftmargin=1.5em,language=C]{Name}
int N = n/t;
#pragma omp parallel
{
  int t_id = omp_get_thread_num();
  for(int k = 0; k < ntimes; k++) {
    for (int i = 0; i < N; i++){
      A[t_id][i] = B[t_id][i] + scalar * C[t_id][i]; 
    }
  }
}
\end{lstlisting}
\end{figure}
\subsection*{Overhead of shared data spaces}
The shape of the curve in the performance results on the triad benchmark is disconcerting. Specifically, bandwidth in L1 is less 
than that in L2.
There is a significant amount of overhead to utilize shared memory parallel applications. We explore the resultant performance bottleneck  with two variants of the triad benchmark: unified data spaces and independent data spaces.

The first variant is implemented with unified data spaces using OpenMP's work sharing constructs. Listing \ref{lst:omp_schedule} is a part of the benchmark driver generated from the \textit{unified data spaces} template with the macro \texttt{CLAUSE} in \texttt{triad.h} set to \texttt{schedule(static, n/t) nowait}.

The second benchmark uses the \textit{independent data spaces} template implemented with distinct data spaces independent of the threads (listing \ref{lst:independent_DS}). The only change needed in the benchmark specification 
is done in the data mapping in the header file. The listing shows the result after macro expansion.

\begin{figure}[tbh]
  \centering
  \includegraphics[width=1\columnwidth]{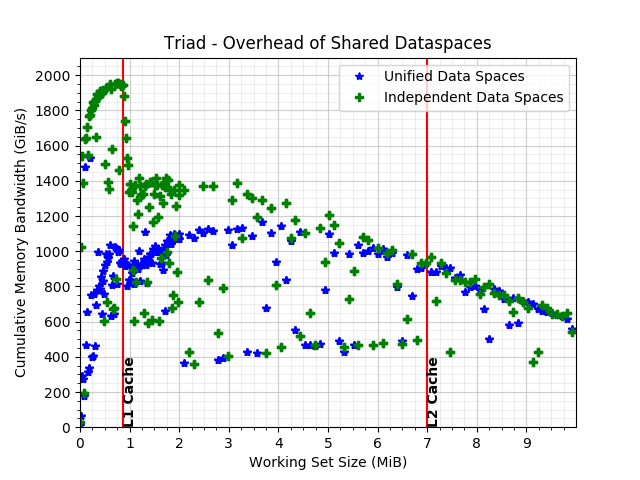}
  \caption{Illustrating the overhead associated with data shared among threads.\label{fig:benchmark_variants}}
\end{figure}

Memory bandwidth results in Figure \ref{fig:benchmark_variants} clearly indicate the benefit of using distinct data spaces over the shared data spaces variant implemented using OpenMP work-sharing and scheduling constructs. Using independent data spaces separates data domains into separate memory regions, eliminating cross-thread communication. This in turn eliminates performance bottlenecks, for example, avoiding multiple threads accessing the same cache line. We observe an approximate two-fold performance boost in the L1 cache with this approach compared to unified data spaces using OpenMP work-sharing constructs, which is deemed to be  efficient.  

\begin{figure}[tbh] 
  \includegraphics[width=1.0\columnwidth]{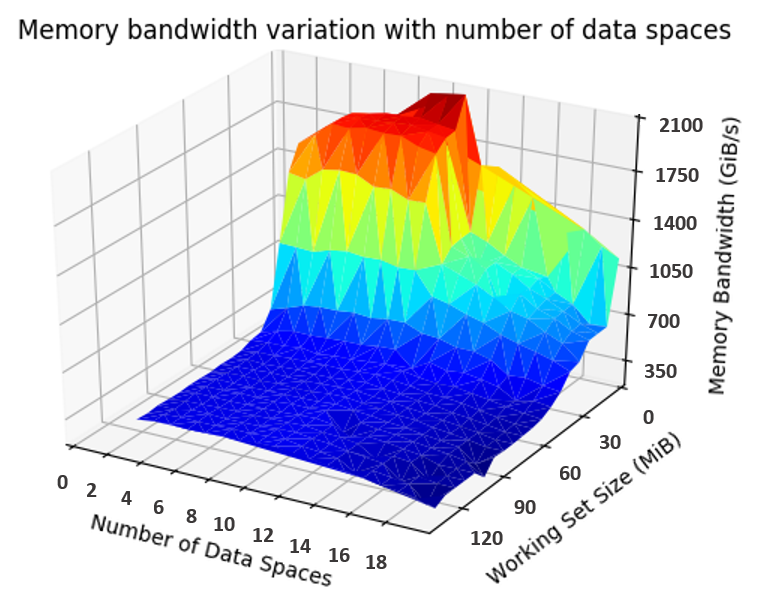}
  \caption{An experiment to identify the number of data streams fetching simultaneously that gives optimal performance on parallel execution with 28 threads.\label{fig:dataspaces}}
\end{figure}
\begin{figure}[tbh]
\begin{lstlisting}[firstnumber=1,caption={Customized benchmark driver with unified spaces illustrating interleaved optimization for triad},label={lst:interleaved},xleftmargin=2em,frame=single,framexleftmargin=1.5em,language=C]{Name}
for (int i = 0; i < n/2; i++){
   A[i] = B[i] + scalar * C[i];
   A[i+n/2] = B[i+n/2] + scalar * C[i+n/2];
}
\end{lstlisting}
\end{figure}

\begin{figure}[h] 
\centering
  \includegraphics[width=0.8\columnwidth]{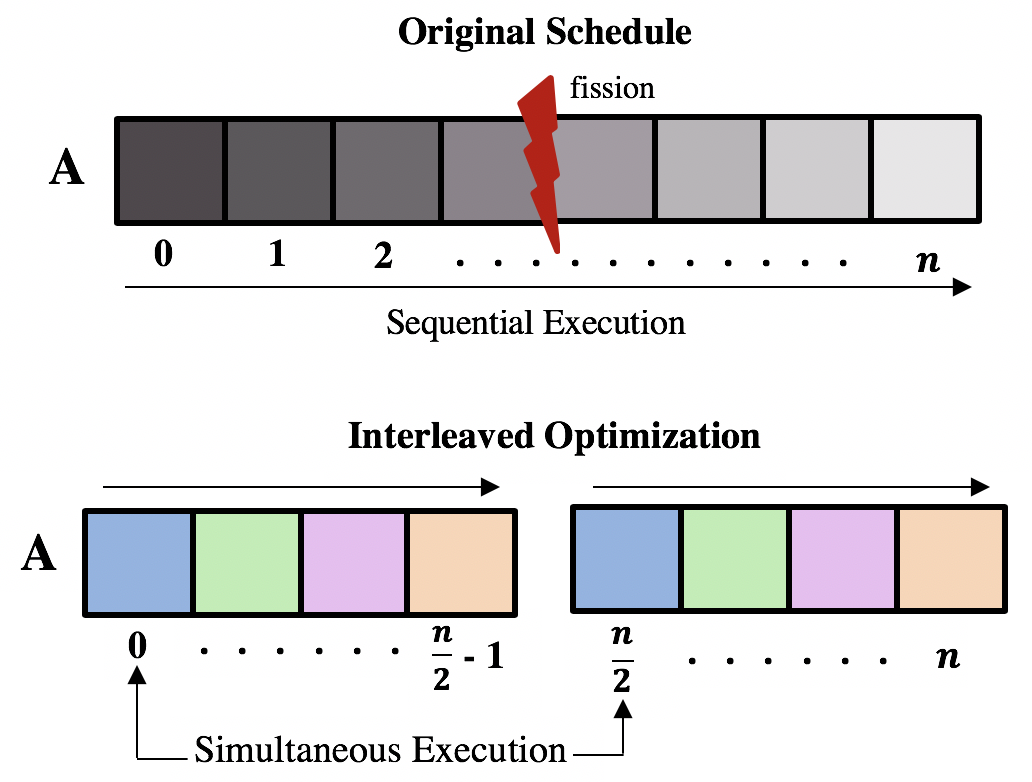}
  \caption{Illustration of interleaved optimization with a single data space of size \textit{n}. \label{fig:Interleaved_illustration}}
\end{figure}

\subsection*{Scheduling to Maximize Bandwidth}
The triad pattern that comprises three data spaces is often considered to yield optimal performance in a given architecture. With the configurability offered by our benchmarking framework, we expand the number of data spaces evaluated from 3 (in triad) to 20 data streams that are simultaneously read in the body of the loop. This is achieved by modifying the statement definition and memory allocation specifications in the header file.

Figure \ref{fig:dataspaces} shows the results of running this experiment in parallel with 28 threads. 
The memory bandwidth values are inconsistent for working set sizes that sit in L1 cache since small data sets are
shared among a large number of threads. 
Considering working sets in L2 cache, where the performance is more consistent, we observe that the achieved memory
bandwidth peaks for 11 data spaces, which is considerably higher when compared to triad that comprises 3 data 
streams. This experiment led to reschedule the execution to triad.

Listing \ref{lst:interleaved} describes the interleaved optimization implemented for triad. This splits each data spaces of size $n$ into two independent blocks of size $\frac{n}{2}$ each. Each of these blocks are fused together to execute in a single iteration and elements in both of these blocks are accessed simultaneously. So, instead of reading three data spaces at the same time, six data streams are accessed concurrently, hence better utilizing the available prefetching lines. Figure \ref{fig:Interleaved_illustration} illustrates how a single data space is interleaved into two blocks and are fused together to be accessed simultaneously within a single iteration.

\begin{figure}[tbh] 
  \includegraphics[width=1.0\columnwidth]{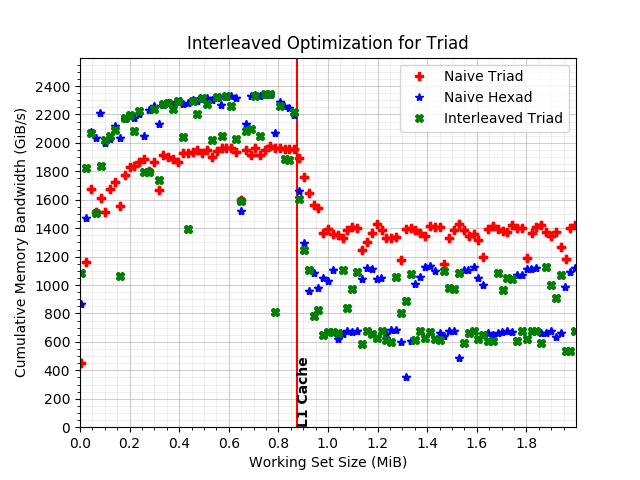}
  \caption{ Interleaved optimization for triad is beneficial in L1 cache on parallel execution with 28 threads.\label{fig:interleaved_triad}}
\end{figure}

Performance results in Figure \ref{fig:interleaved_triad} illustrate the improvement in achieved bandwidth for triad in the L1 cache. 
A significant speedup is observed from the na\"ive triad operation implemented with independent data spaces. For working set sizes falling out of the L1 cache, this optimization is not effective due to poor prefetching.
This further validates the experimental results from Figure \ref{fig:dataspaces}, wherein we achieve higher performance with 6 data spaces (i.e., the na\"ive hexad operation) than 3, which is the case for triad. We attempted interleaving data spaces for triad with interleaving factors greater than two, but we obtain the highest performance when interleaved by 2, due to access to a single cache line exhibiting truly independent data spaces.

\begin{figure*}[thb]
  \centering 
  \subfigure[Number of L1 data cache misses]{\includegraphics[scale=0.54]{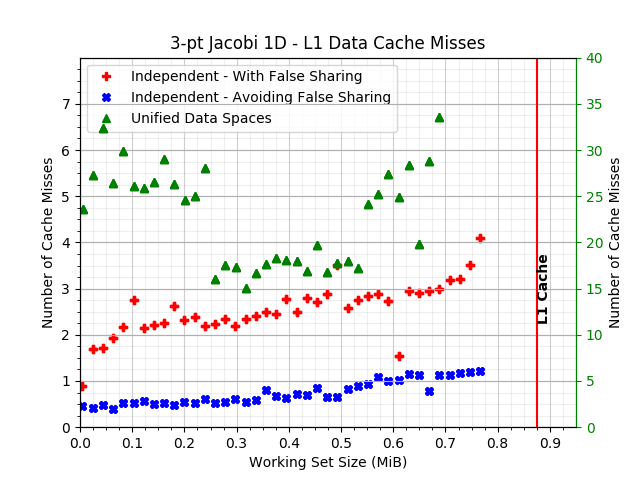}\label{fig:papi_J1D_indDS_L1_DCM}}\quad
  \subfigure[Number of requests to shared cache line]{\includegraphics[scale=0.54]{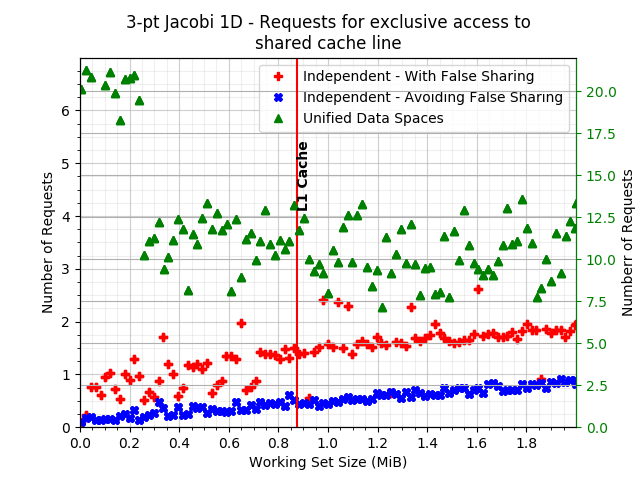}\label{fig:/papi_J1D_indDS_CA_SHR}}\quad
  \caption{Cache misses and cache line requests for 3-pt Jacobi 1D. Measurements for the unified data spaces are plotted along the secondary y-axis for better readability of results.}
  \label{fig:papi_J1D} 
\end{figure*}

\begin{figure}[tbh] 
  \includegraphics[width=1.0\columnwidth]{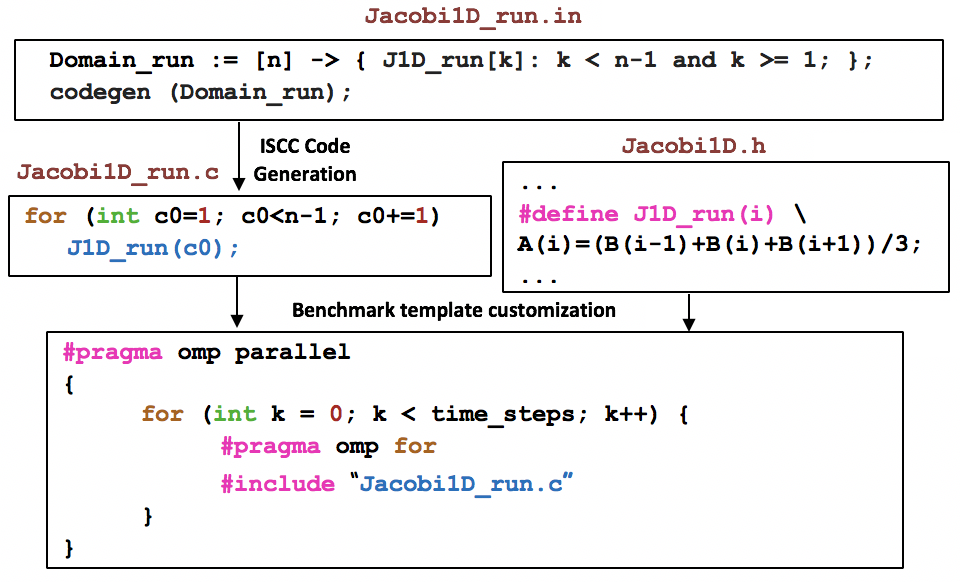}
  \caption{Illustration of custom benchmark generation for 3-pt Jacobi 1D kernel with unified data spaces using the polyhedral model.\label{fig:J1D_demo}}
\end{figure}
\subsection{Multidimensional Jacobi patterns}
Iterative Jacobi stencils are at the core of a wide range of scientific applications and are represented in the Structured Grid motif~\cite{colella2004defining}. These patterns involve nearest neighborhood computations in which each point in a multidimensional grid is iteratively updated by a subset of its neighbors.
The polyhedral model is used to generate benchmark drivers for the Jacobi patterns, as it is helpful to test potential optimizations such as tiling, exercising the flexibility of AdaptMemBench. 
\begin{figure}[tbh]
\begin{lstlisting}[firstnumber=1,caption={The resultant independent data spaces benchmark driver reflecting array padding for Jacobi 1D},label={lst:padding_J1D},xleftmargin=2em,frame=single,framexleftmargin=1.5em,language=C]{Name}
#pragma omp parallel
{
  int t_id = omp_get_thread_num();
  for(int k = 0; k < ntimes; k++) {
     for (int i = 1; i < n - 1; i++){
        A[t_id * 8][i] = (B[t_id * 8][i - 1] + B[t_id * 8][i] + B[t_id * 8][i + 1]) * 0.33; 
     }
  }
}
\end{lstlisting}
\end{figure}

\begin{figure}[tbh] 
  \includegraphics[width=1.0\columnwidth]{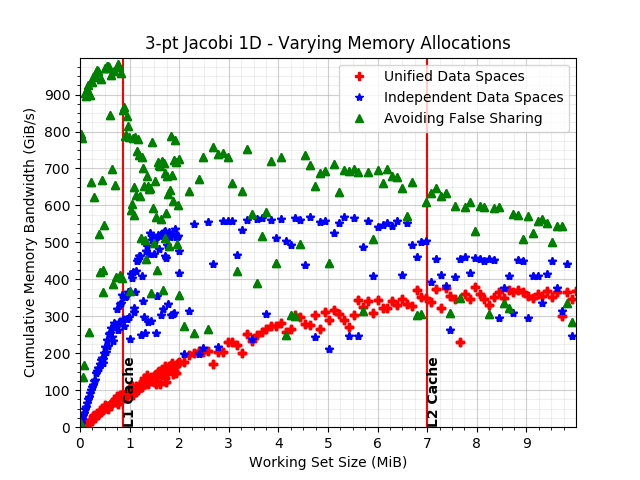}
  \caption{Demonstration of overhead associated with shared data spaces in SMP systems with Jacobi 1D.\label{fig:J1D_indDS}}
\end{figure}
\subsection*{3-pt Jacobi 1D benchmark}
Figure \ref{fig:J1D_demo} demonstrates the process of  custom benchmark generation for this pattern using polyhedral code generation for the input pattern specifications using the \textit{unified data spaces} benchmark template. Allocating independent spaces is advantageous for this pattern as well, as reflected by the memory bandwidth results in Figure~\ref{fig:J1D_indDS}. However, performance scaling in L1 is still an issue, due to false sharing.

\subsection*{Impact of false sharing}

In symmetric multiprocessing systems, where each processor core has dedicated local cache(s), false sharing is a well-known performance issue. False sharing occurs when multiple threads involve in modifying independent variables sharing the same cache line, requiring unnecessary cache flushes and subsequent loads. The potential source of false sharing is multiple threads accessing dynamically allocated or global shared data structures simultaneously.

The impact of false sharing is quantified by recording the performance counters using PAPI. 
We measure the data cache hits in L1 and the requests for exclusive access to shared cache lines in Figure~\ref{fig:papi_J1D_indDS_L1_DCM}. 
We observe that the shared data spaces get affected by cache misses nearly 10 times more than the independent data spaces. 
Please note that Figures~\ref{fig:papi_J1D_indDS_L1_DCM} and~\ref{fig:/papi_J1D_indDS_CA_SHR}  each have a primary and secondary y-axis.
The data plotted using green triangles is associated with the secondary axis (on the right). 
The cache misses recorded for independent data spaces is better, but 
 the variation in number of exclusive requests to clean cache line for the three cases in Figure~\ref{fig:/papi_J1D_indDS_CA_SHR} is much higher for L1 in the case that suffers from false sharing.

Padding arrays is a common solution to overcome false sharing. In the architecture evaluated, each cache line is of size 64 bytes. As shown in Listing~\ref{lst:padding_J1D}, the data spaces of type double are padded with a factor 8 to allocate each element in different cache lines to avoid false sharing. With AdaptMemBench, this can be achieved just by modifying the memory mapping.  Eliminating false sharing leads to a drastic performance speedup in the L1 cache, as the results in Figure~\ref{fig:J1D_indDS} reflect.
The PAPI results were collected by running the same code configurations with a PAPI driver within the framework, and the memory bandwidth results are exclusive of the minimal overhead of accessing hardware counters.

\begin{figure}[t] 
  \includegraphics[width=1.0\columnwidth]{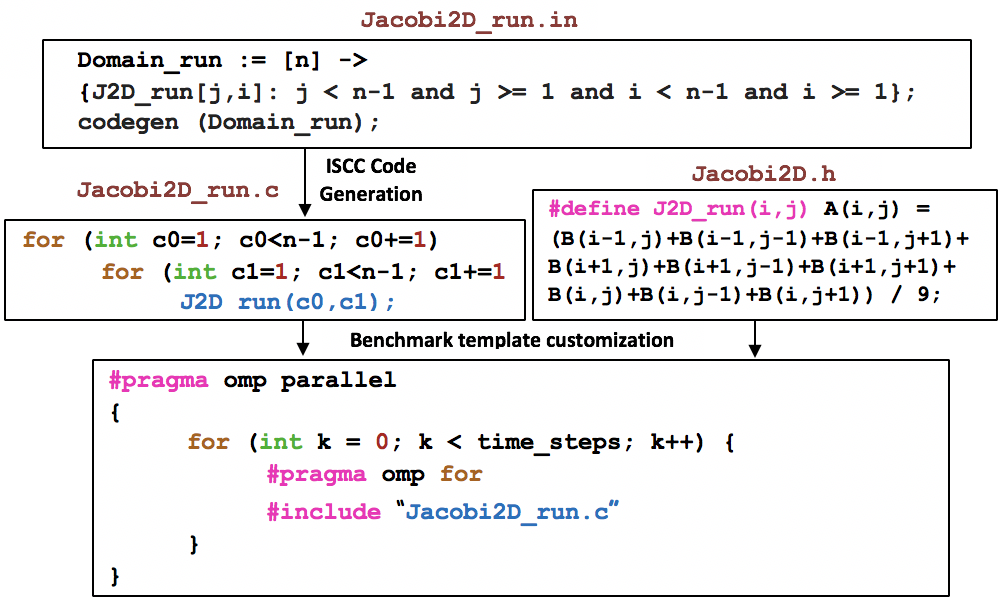}
  \caption{Illustration of custom benchmark generation for 9-pt Jacobi 2D kernel with unified data spaces using the polyhedral model.\label{fig:J2D_demo}}
\end{figure}
\begin{figure}[t] 
  \includegraphics[width=1.0\columnwidth]{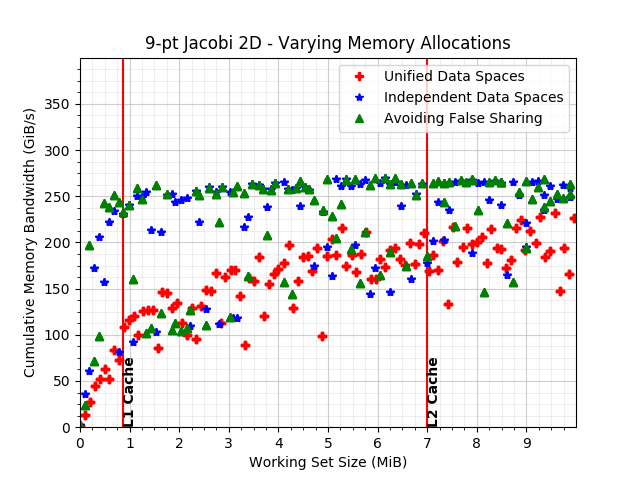}
  \caption{Analyzing the performance bottleneck caused by shared data spaces in Jacobi 2D.\label{fig:J2D_indDS}}
\end{figure}
\begin{figure}[t] 
  \includegraphics[width=1.0\columnwidth]{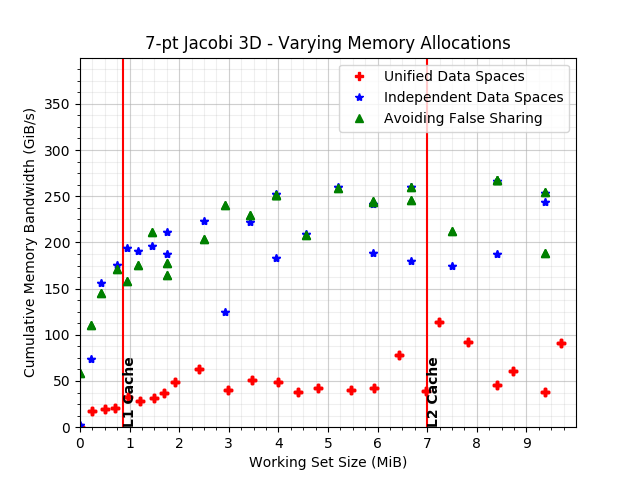}
  \caption{Impact of performance with varying memory allocation in Jacobi 3D.\label{fig:J3D_indDS}}
\end{figure}
\subsection*{Higher dimensional Jacobi patterns}
The process of creating a custom benchmark driver for 9-pt Jacobi 2D using unified data spaces is illustrated in Figure~\ref{fig:J2D_demo}. A 7-point Jacobi 3D  benchmark driver can be similarly created with an added dimension to the code generation script and corresponding modifications to the pattern specification.

From Figures~\ref{fig:J2D_indDS} and \ref{fig:J3D_indDS}, it can be noted that separating data spaces into different memory regions is beneficial for both Jacobi 2D and Jacobi 3D. However, false sharing doesn't affect performance and both the patterns struggle to scale in the L1 cache.

\subsection*{Tiling Optimization for Jacobi transformations}

Rectangular space tiling \cite{irigoin1988supernode} is one of the traditional optimization strategies for stencil computations. Rectangular tiling breaks a large iteration space into a set of smaller iteration spaces, which improves spatial and temporal locality. When  iterating over a large two-dimensional data space applying a multipoint stencil, it is highly probable that one of the neighbors accessed would have fallen out of the cache while the iteration comes around to the same point again. Tiling iteration space eliminates such cache misses and improves data reuse. This optimization is explored, not to provide another data point on the impact of tiling, but to demonstrate the advantages of including polyhedral code representations in the framework.

\subsection*{Tiling Three-dimensional Jacobi}
We implement this spatial tiling strategy on the 7-point Jacobi 3D transformation. The initial approach is to tile in the 3D grid in all directions. Listing \ref{lst:J3D_iscc} shows the ISCC input script and corresponding C code file generated. AdaptMemBench simplifies the testing of this optimization with this input ISCC script as execution schedule file with the other pattern specifications remaining the same as for the na\"ive Jacobi 3D benchmark.

\begin{figure}[tbh]
\begin{lstlisting}[firstnumber=1,label={lst:J3D_iscc},xleftmargin=2em,frame=single,framexleftmargin=1.5em,language=C]{Name}
Domain_run := [n] -> {
    STM_3DS_run[k,j,i] : i <= n and i >= 1 and j<=n and j >= 1 and k<=n and k >= 1;
};
Tiling := [n] -> {
    STM_3DS_run[k,j,i] -> STM_3DS_run[tk,tj,ti,k,j,i]:exists rk,rj,ri:
                    0<=rk<32 and k=tk*32+rk
                and 0<=rj<64 and j=tj*64+rj
                and 0<=ri<16 and i=ti*16+ri;
};
codegen (Tiling * Domain_run);
\end{lstlisting}
\begin{lstlisting}[firstnumber=1,label={lst:J3D_iscc},caption={ISCC script \texttt{Jacobi3D\_xyz\_tiled.in} and the generated C code file \texttt{Jacobi3D\_xyz\_tiled.c}.},xleftmargin=2em,frame=single,framexleftmargin=1.5em,language=C]{Name}
for (int c0 = 0; c0 <= floord(n, 32); c0 += 1)
  for (int c1 = 0; c1 <= n / 64; c1 += 1)
    for (int c2 = 0; c2 <= n / 16; c2 += 1)
      for (int c3 = max(1, 32 * c0);                  c3 <= min(n, 32 * c0 + 31); c3 += 1)
        for (int c4 = max(1, 64 * c1);                  c4 <= min(n, 64 * c1 + 63); c4 += 1)
          for (int c5 = max(1, 16 * c2);                 c5 <= min(n, 16 * c2 + 15); c5 += 1)
            STM_3DS_run(c3, c4, c5);
\end{lstlisting}
\end{figure}

We initially block the iteration space in all the three dimensions, for block sizes $16\times16$, $32\times32$ and $64\times64$. Our results agree with previous experimental evaluation showing no performance gain \cite{kamil2005impact}. 

\begin{figure}[h] 
  \includegraphics[width=1.0\columnwidth]{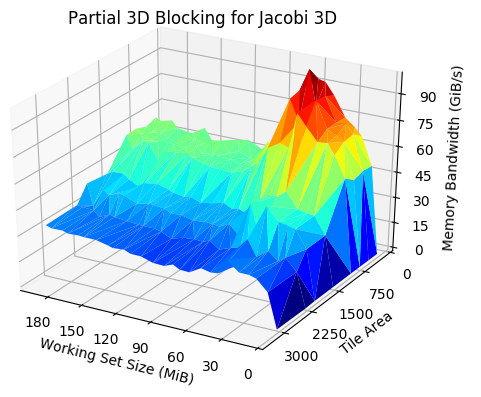}
  \caption{Achieved memory bandwidth with 2D Cache blocking for Jacobi 3D with a tile sweep for sizes ranging from 16 to 64 in both the tiled directions. \label{fig:J3D_partial}}
\end{figure}

We implement the partial blocking strategy \cite{rivera2000tiling} in which blocking is done in two least significant dimensions alone. This results in a series of 2D slices that are stacked one over the other in the unblocked dimension. We tested the efficacy of this technique on grid sizes up to 256, with block sizes ranging from 16 to 64 in both directions. This approach too does not offer any speedup if we compare the peak bandwidth from Figure~\ref{fig:J3D_indDS} with the most performant block area in figure \ref{fig:J3D_partial}. Large on-chip caches affect cache reuse  and thus provide no performance gain with this blocking strategy. Increasing grid sizes would be impractical since many scientific applications, such as computation fluid dynamics, typically use a box size of $64^3$ or less~\cite{adams2015chombo}.

These results confirm conclusions from previous studies \cite{kamil2005impact,datta2009optimization,kamil2006implicit} on these tiling strategies performed for serial execution. We extend these studies to parallel applications and systems using with the flexibility of the polyhedral model offered by AdaptMemBench. Several temporal tiling strategies~\cite{bandishti2012tiling,krishnamoorthy2007effective,wonnacott2000using,frigo2005cache} have proved to be effective for higher dimensional stencil patterns, which are not evaluated in this paper, but the framework can accommodate them.

\section{Related Work}
Several categories of memory benchmarks have been developed over the years. Most relevant to our work are the streaming bandwidth benchmarks, which use a predefined set of access patterns to measure achieved memory bandwidth, and the stencil benchmarks. The following section presents representatives from each benchmarking category.

Our benchmarking framework adds capabilities beyond these benchmarks by offering configurability to explore the performance of scientific applications. It emulates application-specific memory access patterns using the mechanism of polyhedral code generation. It is a flexible and consistent testbed for evaluating various code optimizations without needing to port or modify the entire application.

\subsection{Streaming Bandwidth Benchmarks}
STREAM \cite{mccalpin1995stream} is a microbenchmark that measures sustainable memory bandwidth and the corresponding computation rates for the performance evaluation of high performance computing systems. STREAM measures the performance of four operations: COPY (\texttt{a[i] = b[i]}, measures data transfer without arithmetic), SCALE (\texttt{a[i] = q*b[i]}, with a simple arithmetic operation), SUM (\texttt{a[i] = b[i] + c[i]}, tests multiple load and store operations) and TRIAD (\texttt{a[i] = b[i] + q*c[i]}). The STREAM benchmark does not measure memory bandwidth for small data sizes in the higher levels of memory hierarchy, i.e., in level 1 cache and some portions of level 2 cache, depending on the target architecture. AdaptMemBench calculates the cumulative computation time for the overall execution of the kernel and enabling it to explore achieved performance in higher levels of cache.

MultiMAPS \cite{snavely2002framework} is a benchmark probe designed to measure platform-specific bandwidths, similar to STREAM, it accesses data arrays repeatedly. In MultiMAPS, the access pattern is varied in stride and array size varying spatial and temporal locality. It measures achieved memory bandwidth of different memory levels, different size working sets and a small set of access patterns. This benchmark is most closely related to ours. The primary difference is the ability to include arbitrary memory access patterns, and test optimization strategies.

Stanza triad \cite{kamil2005impact}, a microbenchmark, is a derivative of STREAM, which measures the impact of prefetching on modern microprocessors. It works by comparing the bandwidth measurements by varying stanza length $L$ and stride of access $S$ for different data sizes and predicts performance. This being a serial benchmark, cannot be scaled to parallel applications, and cannot be configured for patterns other than triad.

\subsection{Synthetic memory benchmarks}
Apex-MAP \cite{strohmaier2005apexlatest} is a synthetic benchmark that characterizes application performance, implemented sequentially \cite{strohmaier2004architecture},
 and in parallel using MPI \cite{strohmaier2005apex}.
 This benchmark \textit{approximates} the memory access performance based on concurrent address streams considering regularity of access pattern, spatial locality, and temporal reuse. Using a set of characteristic performance factors, its execution profile is tuned such that these factors act as a proxy for the performance behavior of code with similar characteristics. 

Stencil Probe \cite{kamil2005impact} is a lightweight, flexible stencil application-specific benchmark that explores the behavior of grid-based computations. Stencil Probe mimics the kernels of applications that use stencils on regular grids by modifying the operations in the inner loop of the benchmark. Similar to Stanza Triad, this benchmark is serial and cannot be extended to large-scale parallel applications and systems. Furthermore, this probe is not friendly for testing code optimizations and requires rewriting of the entire the benchmark code for each transformation.  

Bandwidth \cite{smith2008bandwidth} is an artificial benchmark to measure memory bandwidth on x86 and x86\_64 based architectures. This benchmark can be used to evaluate the performance of the memory subsystem, the bus architecture, the cache architecture and the processor. Memory bandwidth is measured by performing sequential and random reads and writes of varying sizes across the levels of the memory hierarchy. However, this benchmark is neither application-specific nor customizable. It measures performance based on a predefined set of memory access patterns and cannot be configured specifically to a target application. Moreover, this benchmark executes serially and cannot be scaled to parallel systems and applications.

\subsection{Application Benchmarks}
Application Benchmarks are used as exemplars of application patterns.
The NAS Parallel Benchmarks \cite{bailey1991parallel} comprises benchmarks developed to represent the major types of computations performed by highly parallel supercomputers and mimic the computation and data movement characteristics of scientific applications. It consists of five \textit{parallel kernel} benchmarks (EP - an embarrassingly parallel kernel, MG - a simplified multigrid kernel, CG - a conjugate gradient method, FT - fast Fourier transforms and IS - a large integer sort) and three \textit{simulated application} benchmarks (LU - lower and upper triangular system solution, SP - scalar pentadiagonal solver and BT - set of block  tridiagonal equations). 

The HPC Challenge benchmark suite \cite{luszczek2005introduction} provides a set of benchmarks that define the performance boundaries of future Petascale computing systems. This hybrid benchmark suite examines the performance of HPC architectures as a function of memory access characteristics using different access patterns. It is composed of well-known computational kernels such as STREAM, HPL \cite{dongarra2003linpack}, matrix multiply, parallel matrix transpose, FFT, RandomAccess and bandwidth/latency tests that span high and low spatial and temporal locality space.

\section{Conclusions}
This paper presents a configurable benchmark framework that captures application-specific memory access patterns that can be expressed using the polyhedral model.
The use of the polyhedral model and associated code generation tools allows for quick development and experimentation with
optimization strategies.
The AdaptMembench framework was used to demonstrate the benefit of using distinct data spaces on threads and the overhead
of OpenMP constructs and false sharing when targeting the L1 cache.

\bibliographystyle{plain}
\bibliography{ref}

\begin{thebibliography}{10}

\bibitem{adams2015chombo}
M~Adams, P~O Schwartz, H~Johansen, P~Colella, T~J Ligocki, D~Martin, ND~Keen,
  Dan Graves, D~Modiano, Brian Van~Straalen, et~al.
\newblock Chombo software package for amr applications-design document.
\newblock Technical report, 2015.

\bibitem{bailey1991parallel}
DH~Bailey, E~Barszcz, JT~Barton, DS~Browning, RL~Carter, L~Dagum, RA~Fatoohi,
  Paul~O Frederickson, Thomas~A L, Rob~S Schreiber, et~al.
\newblock The nas parallel benchmarks.
\newblock {\em The International Journal of Supercomputing Applications},
  5(3):63--73, 1991.

\bibitem{bandishti2012tiling}
V~Bandishti, I~Pananilath, and U~Bondhugula.
\newblock Tiling stencil computations to maximize parallelism.
\newblock In {\em High Performance Computing, Networking, Storage and Analysis
  (SC), 2012 International Conference for}, pages 1--11. IEEE, 2012.

\bibitem{colella2004defining}
P~Colella.
\newblock Defining software requirements for scientific computing.
\newblock 2004.

\bibitem{datta2009optimization}
K~Datta, S~Kamil, S~Williams, L~Oliker, J~Shalf, and K~Yelick.
\newblock Optimization and performance modeling of stencil computations on
  modern microprocessors.
\newblock {\em SIAM review}, 51(1):129--159, 2009.

\bibitem{dongarra2003linpack}
Jack~J Dongarra, Piotr Luszczek, and Antoine Petitet.
\newblock The linpack benchmark: past, present and future.
\newblock {\em Concurrency and Computation: practice and experience},
  15(9):803--820, 2003.

\bibitem{frigo2005cache}
Matteo Frigo and Volker Strumpen.
\newblock Cache oblivious stencil computations.
\newblock In {\em Proceedings of the 19th annual international conference on
  Supercomputing}, pages 361--366. ACM, 2005.

\bibitem{irigoin1988supernode}
Fran{\c{c}}ois Irigoin and Remi Triolet.
\newblock Supernode partitioning.
\newblock In {\em Proceedings of the 15th ACM SIGPLAN-SIGACT symposium on
  Principles of programming languages}, pages 319--329. ACM, 1988.

\bibitem{kamil2006implicit}
Shoaib Kamil, Kaushik Datta, Samuel Williams, Leonid Oliker, John Shalf, and
  Katherine Yelick.
\newblock Implicit and explicit optimizations for stencil computations.
\newblock In {\em Proceedings of the 2006 workshop on Memory system performance
  and correctness}, pages 51--60. ACM, 2006.

\bibitem{kamil2005impact}
Shoaib Kamil, Parry Husbands, Leonid Oliker, John Shalf, and Katherine Yelick.
\newblock Impact of modern memory subsystems on cache optimizations for stencil
  computations.
\newblock In {\em Proceedings of the 2005 workshop on Memory system
  performance}, pages 36--43. ACM, 2005.

\bibitem{kelly1998optimization}
Wayne Kelly.
\newblock Optimization within a unified transformation framework.
\newblock Technical report, 1998.

\bibitem{krishnamoorthy2007effective}
Sriram Krishnamoorthy, Muthu Baskaran, Uday Bondhugula, Jagannathan Ramanujam,
  Atanas Rountev, and Ponnuswamy Sadayappan.
\newblock Effective automatic parallelization of stencil computations.
\newblock In {\em ACM sigplan notices}, volume~42, pages 235--244. ACM, 2007.

\bibitem{luszczek2005introduction}
P~Luszczek, J~J Dongarra, D~Koester, R~Rabenseifner, B~Lucas, J~Kepner,
  J~McCalpin, D~Bailey, and D~Takahashi.
\newblock Introduction to the hpc challenge benchmark suite.
\newblock Technical report, Ernest Orlando Lawrence Berkeley
  NationalLaboratory, Berkeley, CA (US), 2005.

\bibitem{mccalpin1995stream}
John~D. McCalpin.
\newblock Stream: Sustainable memory bandwidth in high performance computers.
\newblock 1991-2007.
\newblock A continually updated technical report.
  http://www.cs.virginia.edu/stream/.

\bibitem{mccalpin1995survey}
John~D McCalpin.
\newblock A survey of memory bandwidth and machine balance in current high
  performance computers.
\newblock {\em IEEE TCCA Newsletter}, 19:25, 1995.

\bibitem{mucci1999papi}
P~J Mucci, S~Browne, C~Deane, and G~Ho.
\newblock Papi: A portable interface to hardware performance counters.
\newblock In {\em Proceedings of the department of defense HPCMP users group
  conference}, volume 710, 1999.

\bibitem{rivera2000tiling}
Gabriel Rivera and Chau-Wen Tseng.
\newblock Tiling optimizations for 3d scientific computations.
\newblock In {\em Proceedings of the 2000 ACM/IEEE conference on
  Supercomputing}, page~32. IEEE Computer Society, 2000.

\bibitem{smith2008bandwidth}
Zack Smith.
\newblock Bandwidth: a memory bandwidth benchmark, 2008.

\bibitem{snavely2002framework}
A~Snavely, L~Carrington, N~Wolter, J~Labarta, R~Badia, and A~Purkayastha.
\newblock A framework for performance modeling and prediction.
\newblock In {\em Supercomputing 2002}, pages 21--21. IEEE, 2002.

\bibitem{strohmaier2005apexlatest}
E.~Strohmaier and Hongzhang Shan.
\newblock Apex-map: A global data access benchmark to analyze hpc systems and
  parallel programming paradigms.
\newblock In {\em Supercomputing, 2005. Proceedings of the ACM/IEEE SC 2005
  Conference}, pages 49--49, Nov 2005.

\bibitem{strohmaier2004architecture}
Erich Strohmaier and Hongzhang Shan.
\newblock Architecture independent performance characterization and
  benchmarking for scientific applications.
\newblock In {\em Modeling, Analysis, and Simulation of Computer and
  Telecommunications Systems, 2004.(MASCOTS 2004). Proceedings. The IEEE
  Computer Society's 12th Annual International Symposium on}, pages 467--474.
  IEEE, 2004.

\bibitem{strohmaier2005apex}
Erich Strohmaier and Hongzhang Shan.
\newblock Apex-map: A synthetic scalable benchmark probe to explore data access
  performance on highly parallel systems.
\newblock In {\em European Conference on Parallel Processing}, pages 114--123.
  Springer, 2005.

\bibitem{verdoolaege2007barvinok}
Sven Verdoolaege.
\newblock barvinok: User guide.
\newblock {\em Version 0.23), Electronically available at http://www. kotnet.
  org/skimo/barvinok}, 2007.

\bibitem{verdoolaege2010isl}
Sven Verdoolaege.
\newblock isl: An integer set library for the polyhedral model.
\newblock In {\em International Congress on Mathematical Software}, pages
  299--302. Springer, 2010.

\bibitem{verdoolaege2012polyhedral}
Sven Verdoolaege and Tobias Grosser.
\newblock Polyhedral extraction tool.
\newblock In {\em Second International Workshop on Polyhedral Compilation
  Techniques (IMPACT’12), Paris, France}, pages 1--16, 2012.

\bibitem{wonnacott2000using}
David Wonnacott.
\newblock Using time skewing to eliminate idle time due to memory bandwidth and
  network limitations.
\newblock In {\em Parallel and Distributed Processing Symposium, 2000. IPDPS
  2000. Proceedings. 14th International}, pages 171--180. IEEE, 2000.

\bibitem{wulf1995hitting}
Wm~A Wulf and Sally~A McKee.
\newblock Hitting the memory wall: implications of the obvious.
\newblock {\em ACM SIGARCH computer architecture news}, 23(1):20--24, 1995.

\end{thebibliography}

\end{document}